\documentclass[conference]{IEEEtran}
\usepackage{graphicx}
\usepackage{balance}

\begin{document}

\makeatletter
\let \@sverbatim \@verbatim
\def \@verbatim {\@sverbatim \verbatimplus}
{\catcode`'=13 \gdef \verbatimplus{\catcode`'=13 \chardef '=13 }} 
\makeatother

\title{Using Assignment Incentives to Reduce Student Procrastination and Encourage Code Review Interactions}

\author{\IEEEauthorblockN{Kevin Wang}
\IEEEauthorblockA{Department of Computer Science\\
University of British Columbia\\
Kelowna, BC, Canada, V1V 2Z3\\
wskksw@mail.ubc.ca}
\and
\IEEEauthorblockN{Ramon Lawrence}
\IEEEauthorblockA{Department of Computer Science\\
University of British Columbia\\
Kelowna, BC, Canada, V1V 2Z3\\
ramon.lawrence@ubc.ca}
}

\maketitle

\begin{abstract}
Procrastination causes student stress, reduced learning and performance, and results in very busy help sessions immediately before deadlines. A key challenge is encouraging students to complete assignments earlier rather than waiting until right before the deadline, so the focus becomes on the learning objectives rather than just meeting deadlines. This work presents an incentive system encouraging students to complete assignments many days before deadlines. Completed assignments are code reviewed by staff for correctness and providing feedback, which results in more student-instructor interactions and may help reduce student use of generative AI. The incentives result in a change in student behavior with 45\% of assignments completed early and 30\% up to 4 days before the deadline. Students receive real-time feedback with no increase in marking time.
\end{abstract}


\section{Introduction}

Assignments are an essential activity for students learning computer science and software engineering to practice programming, design, and analysis. A key challenge is encouraging students to complete assignments in a timely fashion, so they are able to complete an assignment on time without undue stress. Procrastination is a common problem among students \cite{Ellis_Knaus,Laura1984} that leads to lower student performance, mental distress, and many other problems \cite{steelReview,wang2011effects}.  Poor time management often results in students not completing an assignment, handing an assignment in late, frantic use of office hours, or being tempted to have other people or generative AI complete the assignment for them. In all these cases, the learning objectives of the assignment are not achieved as the student is only working towards a deadline.

Ackerman {\em et al.} proposed that incentives for getting an early start, among many other factors, may reduce procrastination \cite{ackerman2005}. Educators incentivized early submission and obtained positive results \cite{Schilling2010,bennett2012}. Felker {\em et al.}  \cite{felker2020impact} implemented an incentivization system for early assignment submission that reduced student cramming. An analysis of student submissions showed that on average students who start working on assignments early perform better \cite{earlybird}, however the last day before the deadline is the most active. 

Our work introduces an easy-to-implement assignment incentive system that offers benefits compared to traditional approaches and other incentive methods.  Students can submit an assignment early for bonus marks with the submission, code review, and marking conducted in office hours or lab time. Synchronous code review with the student has several benefits including allowing the instructor to provide real-time feedback and discussion with students. Often students ignore assignment feedback, especially when the feedback is written and unhelpful \cite{improvetafeedback, jonsson2013review}. Code enhancements and style improvements are more easily provided and received by students when communicating synchronously. Further, the code review process allows the instructor to verify and enhance the student knowledge of the assignment. This provides some level of protection against academic integrity issues, such as code submitted from the Internet or generative AI, as students are challenged to explain their thought process. By conducting the review before the deadline, students have the opportunity to make changes and correct errors.

Evaluation of the technique showed significant student acceptance and behavior change compared to previous course offerings. 31\% of assignments were completed 4 days before the deadline. 92\% of the 44 students who completed the survey claimed that they agree or strongly agree that they are motivated to complete assignments earlier. There was a major shift in office hours utilization with significantly less help sessions the day the assignment was due. Further, many students who did not receive bonus marks benefited, as they started the assignments earlier and thus had more time to learn and complete assignments. Students appreciated the opportunity to demonstrate their work to staff, and the increased engagement with instructors was an additional benefit. The real-time marking resulted in more detailed feedback in approximately the same marking time. 

This paper evaluates if assignment incentivization discourages procrastination, impacts student help seeking and performance, and changes staff grading and feedback practices.
    
\section{Background}

\subsection{Procrastination}

Procrastination has always been a significant problem for students. It is estimated that around 80\% of students procrastinate \cite{Ellis_Knaus}. Researchers have found that reducing procrastination plays a role in increasing student satisfaction \cite{wang2011effects}. A meta-analysis confirmed the prevalence of student procrastination, citing that almost 50\% procrastinate consistently with significant impact on student moods and performance \cite{steelReview}. To counter procrastination, research recommends that students and professors develop more motivation to perform their work and assignments \cite{Jean2022, KachgalRecommendation} and select effective deadlines \cite{deadlineanalysis}.

\subsection{Incentivization}

Incentivization has been shown to be very effective in aiding student learning. Studies demonstrate that bonus marks for mastery of homework correlate with better understanding of class materials \cite{ingalls2018incentivizing}. For example, gamification methods such as providing achievement badges are effective motivators \cite{mobilegameincentive,hakulinen2015,Ibanez2014}. Incentivization provides extrinsic motivation for students, but intrinsic motivation is often preferred for learning and may not also be impacted by incentives.

Educators have explored providing bonus marks as an incentive for submitting assignments early. Bennett {\em et al.} \cite{bennett2012} found that providing bonuses for early assignment submission discourages procrastination and that half of the students would hand in assignments by their early deadline. They concluded that the bonus for early submission can have sanguine effects on student behavior and discourages procrastination while being well received by students \cite{Schilling2010,bennett2012}. Many other educators have found different types and levels of correlation between incentives for early submission and decrease in procrastination across different disciplines \cite{ackerman2005,felker2020impact,winder2023early,crammingBehavior}. 

\subsection{Feedback}

Feedback is important for students. However, there are difficulties scaling \cite{essentialfeedback} feedback. Students can fail to utilize feedback, especially written and unspecific feedback \cite{jonsson2013review}. Research has focused on improving feedback quality \cite{improvetafeedback} and developing automatic, personalized, and context-aware feedback \cite{contextawarefeedback}.   Providing automated code review and quality feedback before assignment due dates can also improve student time management \cite{codequalitygame,dennyIncentive,personalprof}.

\subsection{Instructional Environment}

The study was conducted in an upper-level elective computer science course in the area of database systems. The course had 99 undergraduate students and 8 graduate students. All students are third year or above, and the majority of students in the course were computer science majors. The instructor offered a two hour hybrid office hour session for helping students both in-person and online. Teaching assistants provided a two hour virtual help session from 2 to 4 p.m. every weekday using the HelpMe office hours system \cite{helpme}.

The course had weekly software development assignments that students were expected to complete on their own time. The assignments consisted of writing Java code to interact with databases. Starter code was provided to students consisting of multiple code files, and required the students write between 100 and 300 lines of code. Assignments were done in pairs and expected to take between 6 and 8 hours. JUnit tests were used to determine correctness. Assignments were released at least a week before the assignment start date, and all course material required for the assignment would have been covered on or before the Monday the assignment started. Assignments were intended to be done during the week when assigned but always were due the following week on Friday at midnight. The extra time allows flexibility for students to complete the assignments. There were 10 assignments in total. The assignments and instructions are available\footnote{\scriptsize{\texttt{github.com/rlawrenc/cosc\_404/tree/404\_2023\_Jan/labs}}}.

Previously, the course had a scheduled computer lab time for TA assignment support that was removed due to low student participation and replaced with on-demand virtual office hours. Students did not attend labs when they could work on their own computers. However, this also resulted in limited contact with students and more questions via email. Students would often start an assignment a few days before it was due. This caused problems as the lecture material had already moved on to another topic, students would run out of time to complete an assignment, and last minute questions could not be efficiently answered. On-demand, virtual office hours \cite{helpme} were added to get more student engagement and also to deal with the trend of a high percentage of students starting the assignments late.

\section{Assignment Incentivization Approach}

We use the term {\em assignment incentivization} to describe the practice of encouraging students to start assignments early by offering bonus marks for early completion. A unique approach is to require code review during office hours rather than standard submission or automated grading. Students were informed that they had two ways of submitting their assignments for feedback:

\begin{itemize}

\item {\bf Standard way:} Complete unit tests and submit code files on university learning management system (LMS) before Friday deadline. Feedback and grade provided within a week on the LMS.

\item {\bf Incentive way:} Attend a (virtual) office hours session for code review and grading. If all tests and code review pass on or before Monday, receive 10\% bonus. A 5\% bonus was awarded for completion on or before Wednesday. Bonus marks were only given if student attended help session for code review by instructional staff. No bonus marks for submitting early on the LMS. Students could request marking and code review for no bonus after Wednesday and before the due date.
\end{itemize}

The core idea is that assignment incentive bonus marks were possible for early completion, but instructional staff must verify this completion. Awarding bonus for early completion is not new, but the combination with performing code reviews is unique and especially useful. During the code reviews, instructional staff would provide feedback on code issues, syntax, and style, and ask questions about why students used particular techniques or approaches. This provided useful feedback for the students and allowed instructional staff to verify that the students understood the assignment material and completed it themselves. It also modeled the practice of code reviews for students.

The actual bonus marks and deadlines are flexible for individual courses. The bonus marks were chosen to be enough to make students notice, and the days were chosen so that the assignments would be completed 4 days (Monday) or 2 days (Wednesday) before the due date. Note that if the assignment was started on Monday in one week, the Monday bonus deadline is the following Monday (7 days later) and the final submission deadline was Friday (11 days later). This was strategically chosen so students had some flexibility in submission time while also encouraging timely completion.

Since the assignment marking was based on unit testing, instructional staff could focus on higher-level concepts and techniques rather than verifying correctness during the code reviews. This has the potential to save grading time instead of downloading, running, and evaluating code and providing electronic feedback. A staff member can see the completed tests directly and review code shared by the student in real-time to provide feedback. This approach allows many of the benefits of in-person labs without the cost and scheduling challenges. Students can receive feedback on-demand on their own schedules.

\section{Methodology}

This research uses both survey and quantitative data collected and processed in accordance with a university approved ethics study. Data on student interactions and assignment completion was collected from an online queue system, which provides details regarding instructor-student interactions including wait and help session times, types of questions, time of day when questions were asked, and user information. The second set of quantitative data came from the exam and overall grades of consenting participants.

Of the 107 registered students, 67 students (62\%) consented to analysis of their grades in connection with their help sessions. Help session data collected was analyzed for students that used the HelpMe queue system \cite{helpme} (N=83). The calculation of the help session times was based on the question opening and closing times. Times less than 30 seconds are eliminated as these often occur where no help was given. 

A survey was provided to the students in the last two weeks of class regarding their opinions on  assignment incentivization. There were 44 responses for a 41\% response rate. Interviews with the instructional staff were also conducted.

\section{Results}

\subsection{Discouraging Procrastination}

Table \ref{tab:incentive} contains student answers to survey questions. The majority of the survey asked students about their preference for virtual office hours compared to physical labs. There was very strong support for the flexibility and convenience of virtual office hours. For the incentivization, students reported that they were motivated to finish assignments earlier and attend help sessions more frequently. This is backed up by the fact that 234 of the total 523 assignment submissions (45\%)  were done in help sessions. Among the 234 early submissions, 70\% were graded on Monday, 7\% on Tuesday, 14\% on Wednesday, 7\% on Thursday, and 2\% on Friday. 

The high number of early submissions indicates a change in student behavior and earlier assignment completion from previous course offerings. The majority of the early submissions are completed as early as 4 days ahead of the deadline. Such student behavior suggests that the bonus deadlines are major motivators and  demonstrates the benefits of having a `soft' deadline for receiving bonuses. Free-form student comments in the survey and course reviews were strongly supportive of the incentive system.

\begin{table}[h!]
\centering
\begin{tabular}{|p{4cm}|c|c|c|c|c|}
\hline
{\bf Question} & {\bf SA} & {\bf A} & {\bf N} & {\bf D} & {\bf SD} \\
\hline
 Receiving bonus marks for early assignment completion motivated me to finish assignments earlier. & 82\% & 10\% & 8\% & 0\% & 0\% \\
\hline
 I attended office hours more frequently due to the bonus marks.  & 62\% & 18\% & 13\% & 5\% & 2\% \\
\hline
\end{tabular}
\caption{Incentive System Satisfaction: SA=Strongly Agree, A=Agree, N=Neutral, D=Disagree, SD=Strongly Disagree}
\label{tab:incentive}
\end{table}

\subsection{Impact on Student Help Seeking}

\begin{figure}[!ht]
	\centering
	\includegraphics[width=3.2in]{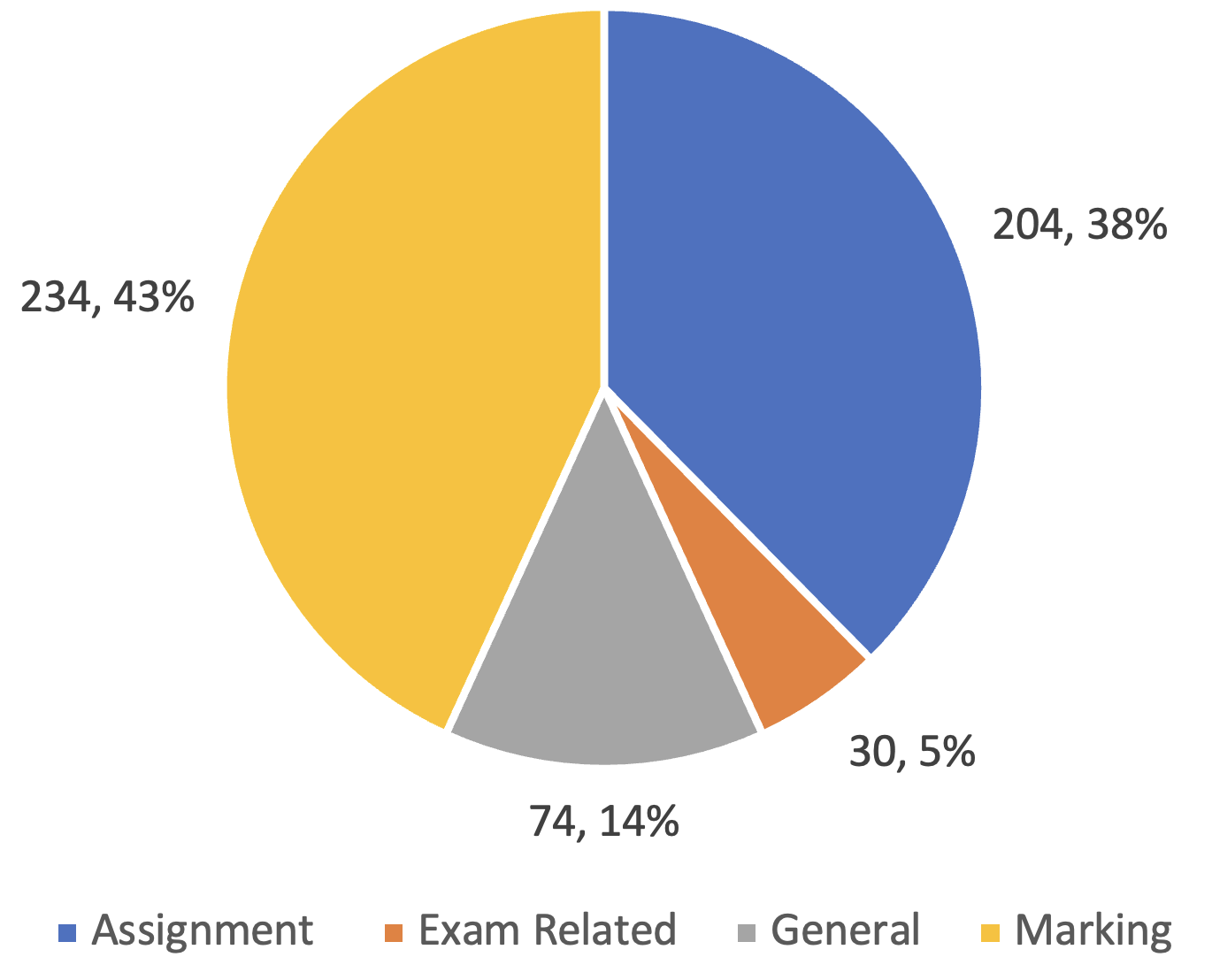}	
        \caption{Question Type Distribution}
        \label{fig:dist}
\end{figure}

The assignment incentivization approach had several positive impacts on student help seeking. First, the number of instructional staff and student interactions increased with 234 interactions out of the total 677 interactions related to assignment marking and evaluation. Figure \ref{fig:dist} has the distribution of questions in all help sessions.

\begin{figure*}[!ht]
	\centering
	\includegraphics[width=6.3in]{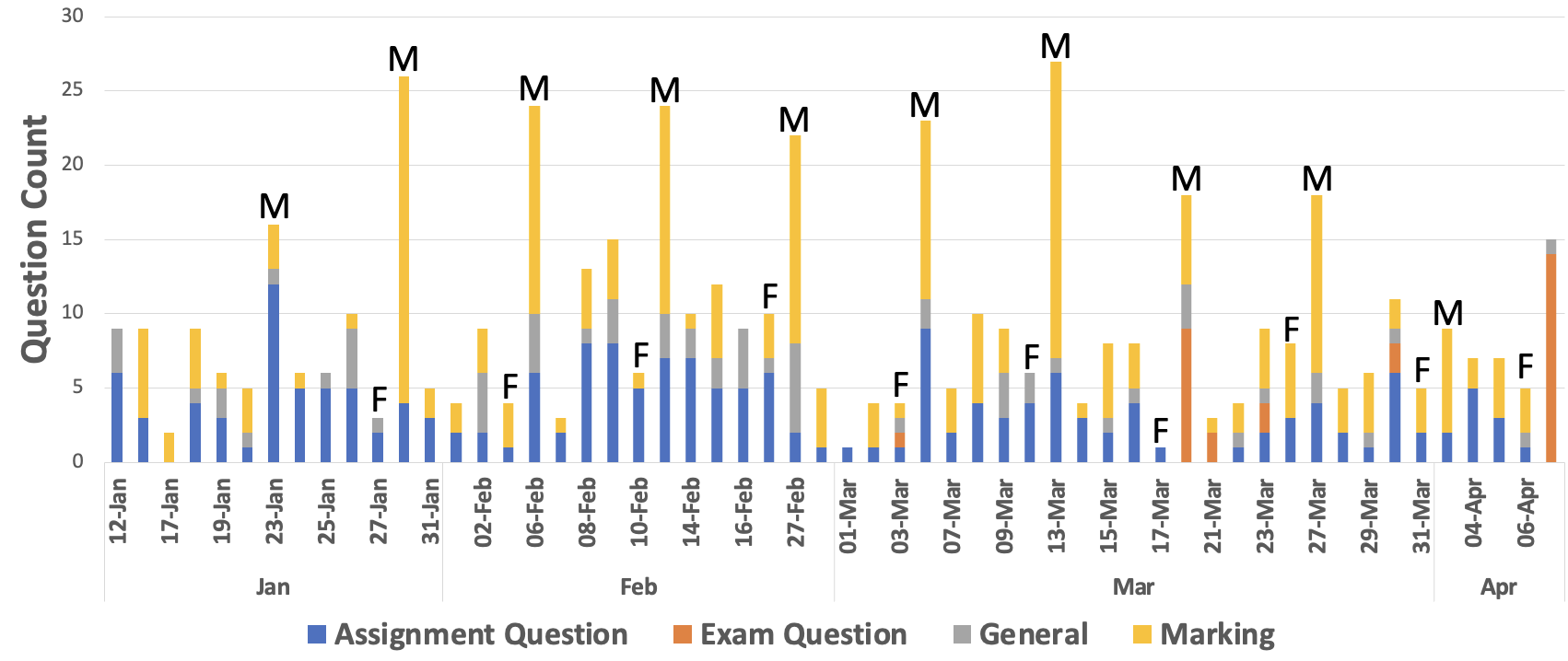}	
        \caption{Help Session Questions by Day}
        \label{fig:util}
\end{figure*}

The incentivization approach also shifted the high-demand times for virtual office hours. Typically, the busiest times for help sessions are within 48 hours before an assignment is due. As shown in Figure \ref{fig:util}, the number of help sessions is small on the Friday ({\bf F}) due date, and instead significant peaks are visible on the Monday ({\bf M}), when the 10\% bonus mark is provided. The incentivization motivated students to complete the assignments earlier. This also allowed more time to help students struggling with the material right before the assignment due date. Furthermore, analyzing student question text demonstrated students used the opportunity to ask follow-on and other questions. 

There are also more general assignment questions on Monday, suggesting that more students may be attending office hours to ask questions on assignments so that they finish them for the 10\% bonus. This corresponds with student data in Table \ref{tab:incentive} showing that students attended office hours more frequently due to the bonus. Assignment incentivization increased the amount of student questions and engagement during office hours. 

\begin{figure}[!ht]
	\centering
	\includegraphics[width=3.3in]{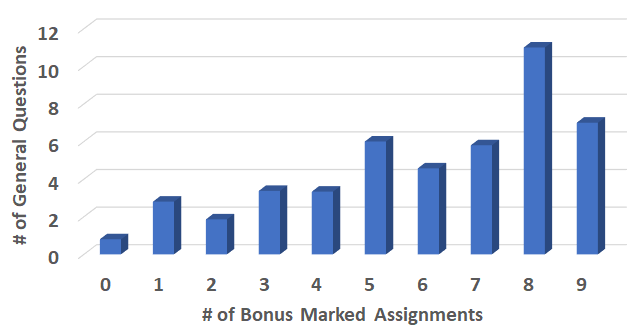}
	\caption{Number of General Questions vs Number of Marked Assignments}
	\label{fig:general}
\end{figure}

In Figure \ref{fig:general} is an analysis of the number of general help sessions per student compared to the number of bonus marked assignments. There is a noticeable improvement in student engagement in general for help session support for students who have used the assignment incentivization. 

\subsection{Instructional Grading and Feedback Practices}

The time for an instructional staff to grade an assignment submitted on the LMS was about 5 minutes. In this time, the staff downloaded the code, executed the tests, and provided a grade and feedback on the LMS submission. Analyzing the submission feedback shows that over 90\% of the submissions do not have any feedback except for the grades. Feedback provided is limited to providing mark breakdowns according to the marking guide when student submissions are only partially correct. There is no higher-level feedback on code style or techniques. This makes sense as TAs must provide grading feedback quickly for a large number of students.

When providing assignment grading during virtual office hours, the time was between 3 and 5 minutes but could extend to 10 minutes or more depending on the code review. Although there is no significant time savings depending on the depth of code review conducted, there is a substantial difference in how the time is spent.

The traditional grading approach spends most of the time on logistics of code evaluation and grade recording and often results in limited feedback to the student (that is often ignored). The synchronous code review spends less than a minute verifying correctness (as tests are pre-executed by student or executed on demand quickly) and the rest of the time providing feedback and review to help the student. This is also an invaluable time to assess student knowledge and encourage academic integrity. The student receives immediate feedback. Overall, there is no time savings, but two TAs were able to support a class of over a hundred students while providing personalized, real-time feedback and verifying academic integrity.

\subsection{Effect on Student Performance}

\begin{figure}[!ht]
	\centering
	\includegraphics[width=3.3in]{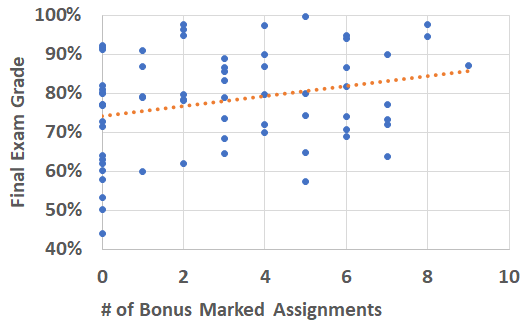}
	\caption{Final Exam vs number of Marked Assignments}
	\label{fig:checked}
\end{figure}

From the queue system data, we obtained the number of times a student requests to mark their assignments early that was compared to their test scores. There is a general improvement in final exam scores for students that received bonus marks due to incentivization (see Figure \ref{fig:checked}).

However, the lower average for students with fewer bonus marked assignments could be attributed to the original ability level and intrinsic motivation of the student, as more capable and motivated students with higher test scores are also more likely to complete assignments early and correctly to obtain bonus marks. A more detailed study would be required to determine if the student learning was positively impacted by the incentivization.

The performance on the final exam for students that had at least one early marked assignment was statistically significant. Overall, the final exam average was 78\% for all students. Students that had no bonus marked assignment had an average of 71\% versus 81\% for students with at least 1 bonus marked assignment.

\section{Discussion}

We found that the assignment incentivization practice of awarding bonus marks for early submission effectively curbs procrastination. The unique approach of assignment marking and code reviews during office hours not only reduced procrastination but also improved feedback and engagement with students. The incentivization boosted questions and student interactions in office hours as captured in the queue system data. We also found that more than 31\% of students completed assignments as early as 4 days before the deadline. There were positive impacts noted by students in terms of changing behavior of attending help sessions and starting assignments.

This research shows that students respond to `soft' deadlines for bonuses that can significantly shift when the effort is being performed on assignments resulting in earlier completion and more students completing assignments on-time. In this case, the vast majority of bonuses are given out on Monday. Having an earlier bonus deadline, irrespective of how many days earlier, can change student behaviors in doing assignments. Knowing this aspect of student behavior, instructors may set bonus deadlines accordingly.

The research was conducted in an upper-level course with well-defined assignments that were verified for correctness using unit tests. The results may be generalizable to other courses as long as assignment correctness can be verified quickly. The technique of grading during help sessions is only possible if assignment correctness is mostly automated and the majority of time is spent engaging with the student during code reviews and questioning. Even for completely automated assessment common in first year courses, performing some synchronous code review with students would allow for increased interactions and more confidence of students completing their own assignments.

Deploying the technique requires flexibility in help session hours. On peak days (i.e., Mondays) multiple staff members would attend help sessions at the start to deal with the surge in demand. This works well for virtual help sessions, but would be slightly more difficult for in-person labs and help sessions. One benefit is that the incentivization allows for better planning on when the highest demand for help sessions will occur and shifts demand from right before the assignment deadline. Setting up such a conducive environment requires a level of coordination and planning.

Assignment incentivization takes help session time that may be required for general assignment help questions. It shifts the time from grading offline to helping students online. If the burden of grading is too high, wait times become high for all students and the approach would not be practical. In the situations when there is over demand for help, assignment incentivization needs to be implemented differently, such as allocating specific time slots for early assignment code review. On the other hand, if there is underutilization of help sessions, this practice can be more beneficial to implement and encourage more student-instructor interactions. 

\section{Future Work and Conclusions}

Assignment incentivization shows great potential as students interact more with instructional staff and complete assignments earlier. There is no specific technology required, and it is easy to deploy as long as assignment correctness is semi-automated. Interactions in virtual office hours have advantages over interactions in class and in labs as they can be more personalized and on-demand to meet student needs. This is especially important in large classes.

Future work will investigate the effects of interactions when grading assignments to see if they help student performance, and if specific questions asked by instructional staff can generate short conversations that lead to learning moments. Determining if code reviews help encourage students to submit their own work rather than using generative AI code is also valuable.

\balance

\bibliographystyle{IEEEtran}
\bibliography{ref.bib}

\end{document}